\documentclass[twocolumn,aps,amsmath,amssymb]{revtex4-1}

\usepackage{hyperref}
\usepackage{multirow}
\usepackage{mathrsfs,amsmath}
\usepackage{verbatim}
\usepackage{amsmath}
\usepackage{amsfonts}
\usepackage{amssymb}
\usepackage{amsthm}
\usepackage{bm}
\usepackage{xparse}
\usepackage{graphicx}
\usepackage{xcolor}
\usepackage{textcase}
\usepackage{url}
\usepackage{lipsum}
\usepackage{appendix}
\usepackage{dsfont}
\usepackage{epstopdf}
\usepackage{footnote}

\usepackage{amsmath}
\usepackage{bbm}
\usepackage{float}
\usepackage{subfigure}
\DeclareDocumentCommand{\Tr}{m m O{\big}}{{\rm Tr}_{\:\!{#1}}#3({#2}#3)}

\begin{document}
\title{One-way network nonlocality of continuous variable entangled networks}
\author{Jun-Li Jiang, Xin-Zhu Liu, Xue Yang} 

\affiliation{School of Information Science and Technology, Southwest Jiaotong University, Chengdu 610031, China}

\author{Xiuyong Ding}
\affiliation{School of Mathematics, Southwest Jiaotong University, Chengdu 610031, China}

\author{Da Zhang}
\email{zhang1556433@sxnu.edu.cn}
\affiliation{School of Physics and Information Engineering, Shanxi Normal University, Taiyuan 030031, China}

\author{Ming-Xing Luo}
\email{mxluo@swjtu.edu.cn}
\affiliation{School of Information Science and Technology, Southwest Jiaotong University, Chengdu 610031, China}

\begin{abstract}
Nonlocality is a key feature of quantum networks and is being studied for its potential applications in quantum communication and computing. Understanding and harnessing nonlocality in quantum networks could lead to the development of faster and more secure communication systems. All the nonclassicalities are limited to discrete variable quantum networks. We propose the first method to verify the network nonlocality of all optical quantum network consisting of two entangled states where one-way classical communication is allowed. This provides the first device-independent method to verify the quantum correlations generated from all optical continuous-variable quantum networks.
\end{abstract}

\maketitle

\section{Introduction}

Bell's theorem highlights the limitations of classical local hidden variable models that depend on shared randomness to account for the probabilistic predictions of quantum theory in all scenarios \cite{Bell,EPR}. The phenomenon known as Bell nonlocality, which is a manifestation of quantumness, can arise from any bipartite entangled system, irrespective of its dimensionality \cite{Gisin1991}. To date, nonlocality has been linked to a deeper understanding of quantum phenomena and plays a significant role in applications within quantum information processing and communication \cite{Gisin1991,Mayers,BCPS,Ekert,Acin2007,Pironio2010}.

The method for detecting bipartite nonlocality can be extended to multipartite scenarios, enabling the characterization of genuine global nonlocality under various model assumptions \cite{Svetlichny,GHZ,Mermin,Bancal}. However, certifying general multipartite entanglements is NP-hard. Another avenue of extension focuses on investigating the correlations that can emerge within network structures \cite{Kimble}. Unlike single entangled systems, these states consist of multiple independent entangled sources, providing a scalable framework for quantum internet and distributed quantum applications \cite{ACL,SSdG,Wehner,Luo2022,Jiang2024}.

The local tensor decomposition of multipartite systems facilitates the exploration of novel quantum features \cite{Yang2022}. However, these network structures introduce new constraints on the investigation of quantumness, particularly because quantum correlations do not form a convex set \cite{Popescu,BGP,Fritz,Chave2015}. This non-convexity complicates the application of linear Bell inequalities, making it challenging to fully characterize these correlations for general networks. Such obstacles not only impede the understanding of quantum features but also drive the definition of new types of quantum nonlocalities \cite{Allen2017,Aberg,networkReview,Renou2021,Abiuso2021}. A promising approach to tackle these challenges involves the use of nonlinear Bell inequalities, especially in tripartite networks engaged in entanglement swapping, which have been successfully generalized to a broader range of network configurations \cite{BRGP,TSCA,BBBC,Tavaki2021}. The objective of these efforts is to certify network nonlocality arising exclusively from networks composed entirely of quantum components \cite{RBBP,Luo2018,networkReview}. Another genuinely nonlocal behaviors \cite{Supic2022,Alej2022,Luo2024} relies on minimal assumptions of classical resources and allows practical applications in quantum information processing on networks.

\begin{figure}[!htb]
\begin{center}
\includegraphics[width=0.46\textwidth]{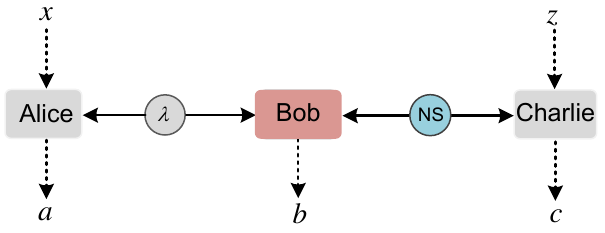}
\end{center}
\caption {Schematic one-way Bell experiment on chain network. One classical source ($\lambda$) send one state to Alice and Bob. One post-quantum no-signaling source (NS) sends one state to Bob and Charlie. Bob performs local operation $B^b$ on his shares with outcome $b$. Alice performs local measurement $A_x$ conditional on the input $x$ on her shares with outcome $a$. Charlie  performs local measurement $C_z$ conditional on the input $z$ on her shares with outcome $c$. Here, we allow Bob send the measurement outcomes $b$ (classical information) to one of other parties.}
\label{fig1}
\end{figure}

In this paper, we aim to explore the network nonlocality of continuous variable (CV) quantum networks. While both CV and discrete variable quantum networks seek to harness quantum resources for information processing, their distinct approaches and capabilities render them suitable for different applications. We first present a Bell-type inequality to witness all correlations generated by hybrid networks containing at least one classical variable and post-quantum sources under one-way classical communication. This inequality provides a one-side device-independent method to verify the network nonlocality stemming from networks composed of two continuous variable entangled states. We propose a comprehensive all-optical experimental design to validate the one-way network nonlocality. This provides the first method to verify the quantum correlations generated from CV entangled quantum networks in a one-side device-independent manner.

\section{One-way network nonlocality of CV chain network}

A multipartite probability distribution is one-way network nonlocal (ONN) relative to a network if and only if it cannot be generated when at least one sources in the network distribute classical physical systems, and the rest are allowed to distribute physical systems only limited by no-signaling. Otherwise, the distribution is one-way network local (ONL). 

\subsection{Hybrid model of chain network}

Consider a chain network consisting of two sources as shown in Fig. \ref{fig1}. There are three different cases for hybrid networks, i.e., two classical variables, one classical variable and one quantum source, and one classical variable and one post-quantum source. Here, we consider the last case which can generate the strongest correlations beyond other two cases. There is one classical source ($\lambda$) that sends one state to Alice and Bob. A post-quantum no-signaling source (NS) sends a state to Bob and Charlie. Bob performs a local measurement $B^b$ on his shares, resulting in outcome $b$. Alice conducts a local measurement $A_x$ based on input $x$, yielding outcome $a$. Charlie performs a local measurement $C_z$ based on input $z$, producing outcome $c$. The joint probability distribution of all outcomes, conditional on inputs, can be decomposed as follows: 
\begin{eqnarray}\label{Eq1}
P(a,b,c|x,z)=\int_{\Omega}p(a|x,\lambda)p(b,c|z,\lambda) d\mu(\lambda)
\end{eqnarray}
where $\mu(\lambda)$ is the probability distribution of classical variable $\lambda$, $p(a|x,\lambda)$ is local response function, and $p(b,c|\lambda,z)$ is a joint response function satisfying the no-signaling principle \cite{PRbox}. To characterize these correlations, we prove the following Bell inequality for any outcome $b$ as 
\begin{eqnarray}
\mathcal{B}=A_0B^bC_0+A_0B^bC_1+A_1B^bC_0-A_1B^bC_1\leq 2 B^b
\label{Eq2}
\end{eqnarray}
 where the correlations are defined by $A_xB^bC_z=\sum_{a,c=\pm1}acP(a,b,c|x,z)$. This inequality has been proved to verify the full network nonlocality of discrete-variable entangled networks \cite{Luo2024} with maximal violation as $2\sqrt{2}B^b$, where one-way classical communication is not allowed. 

To witness the CV quantum networks with all optical experiments, we assume one-way classical communication is  allowed. We modify the correlator conditional on any measurement outcome $b$ as follows
\begin{eqnarray}
\langle A_xC_z\rangle &=&\langle A_x\rangle \langle C_z\rangle, \forall x,z,
\label{Eq2a}\\
\mathcal{B}_{b}&=&(\langle \langle A_0B^bC_0\rangle +\langle A_0B^bC_1\rangle +\langle A_1B^bC_0\rangle 
\nonumber\\
&&-\langle A_1B^bC_1\rangle) /\langle B^b\rangle
\nonumber\\
&=&\langle A_0C_0|B^b\rangle+\langle 
 A_0C_1|B^b\rangle
 \nonumber\\
&&+\langle A_1C_0|B^b\rangle-\langle A_1C_1|B^b\rangle
\nonumber\\
&\leq & 2
\label{Eq2b}
\end{eqnarray}
where $\langle A_xC_z|B^b\rangle$, i.e., the correlation of Alice and Charlie conditional on the outcomes of Bob, is defined by $\langle A_xC_z|B^b\rangle=\sum_{a,c=\pm1}acP(a,c|x,z,b)$, $P(a,c|x,z,b)$ denotes the conditional distribution of outcome $a,c$ conditional on inputs $x$ and $z$ and outcome $b$. The first equality means the independence of Alice and Charlie while the second inequality means that Alice and Charlie cannot create the entanglement correlations conditional on the outcomes of Bob for any hybrid networks. Specially, suppose Alice and Bob share the classical hidden variable. The first equality (\ref{Eq2a}) can be followed by using the independence of Alice and Charlie.  Moreover, the operator $\mathcal{B}_b$ in the left side of Eq.~\eqref{Eq2b} is a linear combination of Alice's measurements, without loss of generality, assume deterministic outputs and have $A_i\in \{\pm 1\}$. By substituting all possibilities for $A_i$, one arrives to
\begin{eqnarray}    \mathcal{B}_b&=&\langle A_0B^bC_0+A_0B^bC_1+A_1B^bC_0-A_1B^bC_1\rangle /\langle B^b\rangle
    \nonumber\\
    &\leq & \max\{|\langle B^bC_0+B^bC_1+B^bC_0-B^bC_1\rangle /\langle B^b\rangle|,
  \nonumber\\  
   && |\langle B^bC_0+B^bC_1-B^bC_0+B^bC_1\rangle /\langle B^b\rangle|
    \}  \nonumber\\
    &\leq & \max\{|\langle C_0+C_1+C_0-C_1\rangle |,
    \nonumber\\
    &&
    |\langle C_0+C_1-C_0+C_1\rangle|  
    \}  \nonumber\\
    &\leq &2,
\end{eqnarray}
where the last bound can be found that the conditional distribution of $P(c|z,b)$ satisfies the normalization. Similar result holds for the case that Bob and Charlie share the classical hidden variable. This means that the inequality \eqref{Eq2b} holds whenever one of the sources distributes classical systems. With this extension, violating the inequality ~\eqref{Eq2b} can verify the quantum correlations arisen from the  tripartite quantum network in Fig.\ref{fig1}. 

\begin{figure}
\begin{center}
\resizebox{240pt}{170pt}{\includegraphics{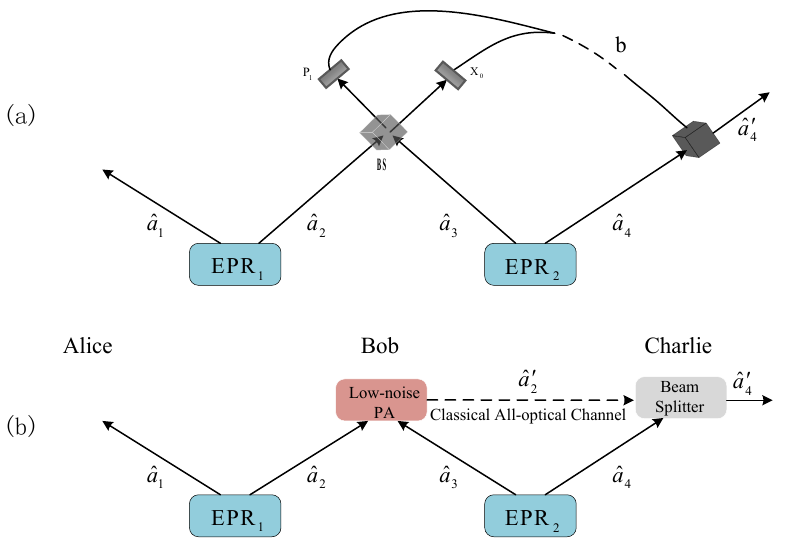}}
\end{center}
\caption{Schematics of entanglement swapping. $\hat{a}_1(\hat{a}_2)$ and $\hat{a}_3(\hat{a}_4)$ represent the annihilation operators for the signal and idler beams of $EPR_1$ and $EPR_2$, respectively. Alice and Charlie begin with two separable particles.  (a) Schematic of electro-optic entanglement swapping. Bob performs a local Bell measurement using a beam splitter and two homodyne detectors, and transmits the measurement outcome $b$ to Charlie, who then performs the corresponding displacement operation based on the measurement result to complete the entanglement swapping. (b) Schematic of AOES. To accomplish AOES, Bob applies a low-noise PA operation to the two modes that he possesses,  and then transmits the amplified state $\hat{a}^{\prime}_2$ to Charlie through an all-optical channel. Charlie couples the received state with $\hat{a}_4$, while Alice does not perform any operations. }
\label{fig2}
\end{figure}

\subsection{CV chain network}

In CV quantum scenarios, the system consists of two continuous-variable entangled sources. In this quantum network, independent particles can be activated to become entangled through local operations on the middle nodes. The concept of entanglement swapping lies at the core of future quantum networks, as it is an essential component of a quantum repeater \cite{Kimble}. This remarkable concept of entanglement swapping has also been extended to the CV regime of quantum information, with the goal of demonstrating its unconditional implementation \cite{16,17}. Our objective is to characterize these experiments in a device-independent manner. Interestingly, for continuous-variable entangled networks, we observe a similar violation of the inequality (\ref{Eq2b}), although there are a large number of measurement outcomes for the Bell measurement. 

In the continuous-variable (CV) quantum information system, implementation typically relies on parametric down-conversion and feed-forward technology for optical field quadrature entanglement \cite{Zukowski}. However, the feed-forward approach, which involves opto-electronic and electro-optic conversions, restricts the entanglement swapping bandwidth. To address this limitation, we employ an all-optical entanglement swapping (AOES) scheme \cite{ref-29} that eliminates the need for measurement, thereby avoiding the conversions commonly used in quantum information protocols \cite{ref-210}. In our scheme, a low-noise parametric amplifier (PA) based on a double-$\Lambda$ configuration four-wave mixing (FWM) \cite{ref-211,ref-213,ref-214} performs the Bell State Measurement (BSM) without detection, while a beam splitter is used to facilitate state displacement.

\subsection{One-way Bell test on all-optical CV entanglement network}

Our discussion in this section on all-optical entanglement swapping is based on the recent experiment \cite{ref-29}, who witnessed the entanglement swapping by evaluating the variance of the amplitude quadrature difference (phase quadrature sum) of the modes on both sides. Instead, we propose a device-independent approach, where each pair of adjacent parties share two CV entangled states to implement AOES using double-$\Lambda$ configuration FWM process in the 85Rb vapor cell. In this double-$\Lambda$ FWM process, two pump photons produce one signal photon and one idler photon. The interaction Hamiltonian for these two FWM processes can be expressed as
\begin{eqnarray}
&\hat{H}_1=i\hbar\gamma_1(\hat{a}^\dag_1\hat{a}^\dag_2-\hat{a}_1\hat{a}_2),\nonumber\\
&\hat{H}_2=i\hbar\gamma_2(\hat{a}^\dag_3\hat{a}^\dag_4-\hat{a}_3\hat{a}_4),
\label{A1}
\end{eqnarray}
where $\hat{a}_1 (\hat{a}_2)$ and $\hat{a}_3 (\hat{a}_4)$ represent the annihilation operators for the signal and idler beams of $EPR_1$ and $EPR_2$, respectively \cite{ref-29}. $\gamma_1(\gamma_2)$ is the interaction strength of the FWM process to generate $EPR_1$ ($EPR_2$), and the intensity gain is $G_1=\cosh^2{\gamma_1 \tau}$ ($G_2=\cosh^2{\gamma_2 \tau}$) for $FWM_1 (FWM_2)$. $\tau$ is the interaction timescale. Based on Eq. \eqref{A1}, the output fields of two FWM processes can be described as
\begin{eqnarray}
&\hat{a}_1=\sqrt{G_1}\hat{a}_0+\sqrt{G_1-1}\hat{v}_1^{\dagger},\nonumber\\
&\hat{a}_2=\sqrt{G_1}\hat{v}_1+\sqrt{G_1-1}\hat{a}_0^{\dagger},\nonumber\\
&\hat{a}_3=\sqrt{G_2}\hat{v}_2+\sqrt{G_2-1}\hat{v}_3^{\dagger},\nonumber\\
&\hat{a}_4=\sqrt{G_2}\hat{v}_3+\sqrt{G_2-1}\hat{v}_2^{\dagger},
\label{C2}
\end{eqnarray}
where $\hat{v}_1$ is the vacuum state for $FWM_1$, whereas $\hat{v}_2$ and $\hat{v}_3$ are the vacuum states for $FWM_2$. By defining the squeezing parameter $r_1=\gamma_1 \tau(r_2=\gamma_2 \tau)$, the above output mode can be described as the Heisenberg representation as 
\begin{eqnarray}
\hat{x}_{a_1}=\frac{1}{\sqrt{2}}(e^{+r_1}\hat{x}_{a_0}+e^{-r_1}\hat{x}_{v_1}),
\nonumber\\
\hat{p}_{a_1}=\frac{1}{\sqrt{2}}(e^{-r_1}\hat{p}_{a_0}+e^{+r_1}\hat{p}_{v_1}),\nonumber\\
\hat{x}_{a_2}=\frac{1}{\sqrt{2}}(e^{+r_1}\hat{x}_{a_0}-e^{-r_1}\hat{x}_{v_1}),\nonumber\\
\hat{p}_{a_2}=\frac{1}{\sqrt{2}}(e^{-r_1}\hat{p}_{a_0}-e^{+r_1}\hat{p}_{v_1}),\nonumber\\
\hat{x}_{a_3}=\frac{1}{\sqrt{2}}(e^{+r_2}\hat{x}_{v_2}+e^{-r_2}\hat{x}_{v_3}),\nonumber\\
\hat{p}_{a_3}=\frac{1}{\sqrt{2}}(e^{-r_2}\hat{p}_{v_2}+e^{+r_2}\hat{p}_{v_3}),
\nonumber\\
 \hat{x}_{a_4}=\frac{1}{\sqrt{2}}(e^{+r_2}\hat{x}_{v_2}-e^{-r_2}\hat{x}_{v_3}),\nonumber\\
\hat{p}_{a_4}=\frac{1}{\sqrt{2}}(e^{-r_2}\hat{p}_{v_2}-e^{+r_2}\hat{p}_{v_3}),
\label{A7}
\end{eqnarray}
where
$r_1$ and $r_2$ correspond to the compression coefficients of the two EPR pairs, respectively. After the generation, Alice receives $\hat{a}_1$ and Charlie receives $\hat{a}_4$, indicating that they each start with distinct, unentangled particles. $\hat{a}_2$ and $\hat{a}_3$ are then sent to Bob. 

\begin{figure}
\begin{center}
\resizebox{240pt}{200pt}{\includegraphics{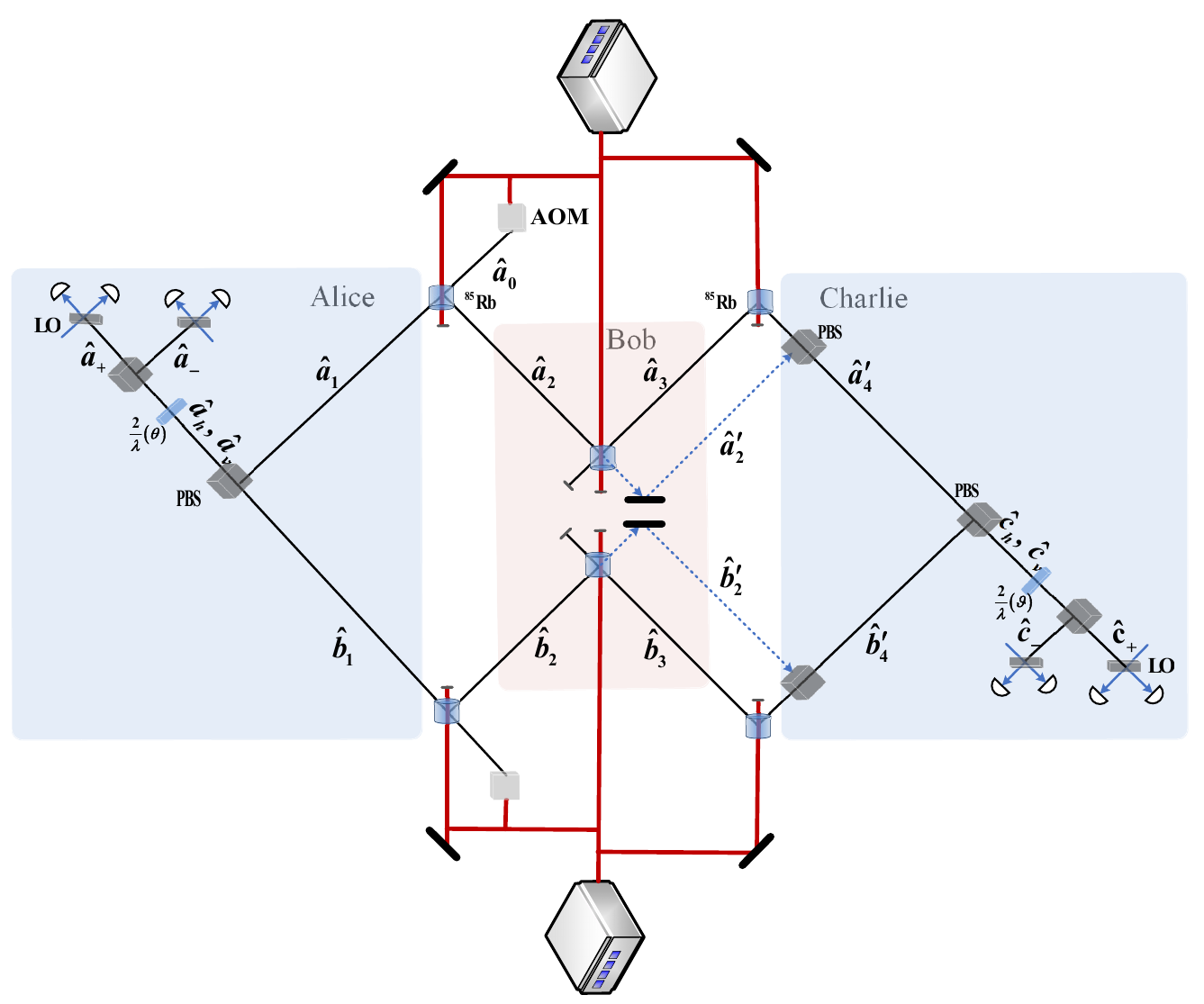}}
\end{center}
\caption{A schematic diagram of Bell-type correlations experiment. PBS, polarization beam splitter; HWP, half wave plate; LO, local oscillator; The HWP and the second PBS enclosed by red dashed rectangle together play the role of adjustable beam splitter. Alice, Bob and Charlie perform two of the above AOES at the same time. After the above AOES process, Alice has beam $\hat{b}_1$ (horizontal polarization) and beam $\hat{a}_1$ (vertical polarization), and then combines them using PBS to form two output modes $\hat{a}_h$ and $\hat{a}_v$. Charlie reorganizes the beams $\hat{a}'{}_4$ and $\hat{b}'{}_4$ in the same way to form two output modes $\hat{c}_h$ and $\hat{c}_v$. Then, Alice and Charlie mix their modes by $\theta$ and $\vartheta$, respectively. The resulting modes, $\hat{a}_+$, $\hat{a}_-$, $\hat{c}_+$, and $\hat{c}_-$, are measured with homodyne detectors. }
\label{fig3}
\end{figure}

To witness the entanglement swapping on CV quantum networks, one method is similar to the experiments on discrete-variable quantum networks \cite{Zukowski}. In the idealized electro-optical scheme depicted in  Fig. \ref{fig2}(a), Bob performs a local Bell measurement using a beam splitter and two homodyne detectors \cite{T2000}. The photonic currents of the two homodyne detectors are given by $X_0=\frac{K}{2\sqrt{2}}(\hat{a}_{2}+\hat{a}^{\dag}_{2}+\hat{a}_{3}+\hat{a}^{\dag}_{3})$ and $P_1=\frac{K}{2\sqrt{2}}(\hat{a}_{2}-\hat{a}^{\dag}_{2}-\hat{a}_{3}+\hat{a}^{\dag}_{3})$ with a  probability density $p(X_0,P_1)$, where $K$ is a proportionality constant. The photonic currents can be combined to form a classical signal as 
\begin{eqnarray}
b'=\sqrt{2}(X_0+iP_1)=K(\hat{a}_2+\hat{a}^{\dag}_3)
\label{eq1}
\end{eqnarray}
with a probability density $p(b')=p(X_0,P_1)$. After received the signal $b'$,  Charlie performs a local displace $\sqrt{2}(X_0+iP_1)$ to shift the local mode. After the two entanglement swapping operations described above, Alice and Charlie implement local measurements to complete the Bell test, following a process similar to the Bell test process described in Fig. \ref{fig3}. In this case, Bob needs to send the classical information of $X_0$ and $P_1$. This feed-forward approach will restrict the entanglement swapping bandwidth \cite{ref-29}. 

Instead, our scheme in what follows is based on a recent experiment that considers a replacement of local Bell measurement with a low-noise PA to realize the information transfer without detection \cite{ref-29}, as shown in Fig. \ref{fig2}(b). Specifically, for a low-noise PA the output will be given by \cite{ref-29}: 
\begin{eqnarray}\label{D1}
\hat{a}'{}_2=\sqrt{G_3}\hat{a}_2+\sqrt{G_3-1}\hat{a}_3^\dagger,
\end{eqnarray}
where $G_3$ is the intensity gain of PA. If the gain $G_3\gg1$, the modes $\hat{a}_1$ and $\hat{a}'{}_2$ shared by Alice and Bob are separable states, see Lemma 1, where we shown the uncertainties of the conjugate quadrature variables $\hat{x}_{a'{}_2}$ and $\hat{p}_{a'{}_2}$ are significantly greater than the quantum limit, i.e., $\bigtriangleup\hat{x}_{a'{}_2}^2,\bigtriangleup\hat{p}_{a'{}_2}^2\gg \frac{1}{4}$  \cite{Ralph1999,Liu2024}. 
This means that simultaneous measurements of the conjugate quadratures can extract all the information carried by $\hat{a}'{}_2$ with a negligible penalty. The quantum noise added owing to the simultaneous measurements will be negligible compared with the amplified quadrature uncertainties. Similarly, $\hat{a}'{}_2$ can suffer propagation loss and compensating linear amplification without degradation. Hence, this can be regarded as an effective classical channel, where $\hat{a}'{}_2$ is treated as a classical field \cite{Ralph1999}. In fact,  Eqs. \eqref{eq1} and \eqref{D1} are identical under the substitution $K=\sqrt{G_3}$ in the limit $G_3\gg1$, where the mode $\hat{a}_2'$ is then regarded as the outcome $b$ in Eq.(\ref{Eq2b}).

\textbf{Lemma 1}. Under the assumption in the present scheme, the modes $\hat{a}_1$ and $\hat{a}_2'$ are separable.

\textbf{Proof}. From Eq. \eqref{D1}, the Heisenberg expression of the mode $\hat{a}_2'$ can be given by
\begin{eqnarray}\label{a1}
\hat{x}'{}_{a_2}=&\sqrt{\frac{G_3}{2}}(e^{+r_1}\hat{x}_{a_0}-e^{-r_1}\hat{x}_{v_1})\nonumber\\&+\sqrt{\frac{G_3-1}{2}}(e^{+r_2}\hat{x}_{v_2}+e^{-r_2}\hat{x}_{v_3}),
\end{eqnarray}
 and
 \begin{eqnarray}
\hat{p}'{}_{a_2}=&\sqrt{\frac{G_3}{2}}(e^{-r_1}\hat{p}_{a_0}-e^{+r_1}\hat{p}_{v_1})\nonumber\\&-\sqrt{\frac{G_3-1}{2}}(e^{-r_2}\hat{p}_{v_2}+e^{+r_2}\hat{p}_{v_3}).
\end{eqnarray}
Combining Eq. \eqref{A7} and Eq. \eqref{a1}, we obtain the correlation matrix of the modes $\hat{a}_1$ and $\hat{a}'{}_2$ as
\begin{eqnarray}
\textrm{Cor}(\hat{a}_1,\hat{a}'{}_2)=\begin{pmatrix}
n_1 & 0 & c_1 & 0 \\
0 & n_2 & 0 & c_2 \\
c_1 & 0 & m_1 & 0 \\
0 & c_2 & 0 & m_2
\end{pmatrix}
\end{eqnarray}
where $n_1=n_2=\frac{1}{4}\cosh{2r_1}$, $m_1=m_2=\frac{1}{4}[G_3\cosh{2r_1}+(G_3-1)\cosh{2r_2}]$, and $c_1=-c_2=\frac{1}{4}\sqrt{G_3}\sinh{2r_1}$.

In what follows, we consider the following type of EPR-like operators:
\begin{eqnarray}\label{a2}
&\hat{u}=a\hat{x}_{a_1}-\frac{c_1}{|c_1|}\frac{1}{a}\hat{x}'{}_{a_2},\nonumber\\
&\hat{v}=a\hat{p}_{a_1}-\frac{c_2}{|c_2|}\frac{1}{a}\hat{p}'{}_{a_2},
\end{eqnarray}
where $a^2=\sqrt{\frac{m_1-\frac{1}{4}}{n_1-\frac{1}{4}}}$. It has shown that the two modes $\hat{a}_1$ and $\hat{a}'{}_2$ are separable if and only if the EPR-like operators satisfy the following inequality \cite{duan}:
\begin{eqnarray}
\label{a3}
\langle\Delta\hat{u}^2\rangle+\langle\Delta\hat{v}^2\rangle
\geq \frac{a^2}{2}+\frac{1}{2a^2}.
\end{eqnarray}

Combining Eqs. \eqref{A7}, \eqref{a1}, and \eqref{a2}, it follows that
\begin{eqnarray}
\label{a4}
&\langle(\Delta\hat{u})^2\rangle+\langle(\Delta\hat{v})^2\rangle
-\frac{a^2}{2}-\frac{1}{2a^2}
\nonumber\\
&=a^2(n_1-1)+\frac{1}{a^2}(m_1-1)-2c_1
\nonumber\\
&=(\cosh{2r_1}-1)(G_3(\cosh{2r_2}-1)-(\cosh{2r_2}+1))
\nonumber\\
&>0
\end{eqnarray}
when $G_3\gg1$. This means the modes $\hat{a}_1$ and $\hat{a}'{}_2$ are separable after Bob's operations. Similarly,  we can prove the modes $\hat{a}'{}_2$ and $\hat{a}_4$ are separable. $\Box$

After amplification,  Bob sends the mode $\hat{a}'{}_2$ to Charlie via an optical channel. Charlie couples  $\hat{a}'{}_2$ to $\hat{a}_4$ and then attenuates the coupled beam with a beam splitter that has a transmission of $\varepsilon=1/G_3$. And then, under the approximation $G_3\gg 1$, $\hat{a}_4$ is translated to $\hat{a}'{}_4$ which can be expressed as
\begin{eqnarray}\label{D2}
\hat{a}'{}_4&=\sqrt{\varepsilon}\hat{a}'{}_2-\sqrt{1-\varepsilon}\hat{a}_4\nonumber\\
&=\frac{1}{\sqrt{G_3}}\hat{a}'{}_2-\sqrt{1-\frac{1}{G_3}}\hat{a}_4.
\end{eqnarray}
By inserting Eq. \eqref{D1} into Eq. \eqref{D2}, we get
\begin{eqnarray}\label{D3}
\hat{a}'{}_4&=\hat{a}_2+\sqrt{\frac{G_3-1}{G_3}}(\hat{a}^\dag_3-\hat{a}_4).
\end{eqnarray}
Then from Eqs. \eqref{C2} we get:
\begin{eqnarray}\label{D31}
\hat{a}'{}_4
&=\hat{a}_2+\sqrt{\frac{G_3-1}{G_3}}[\hat{v}^\dag_2(\sqrt{G_2}-\sqrt{G_2-1})\nonumber\\&\quad -\hat{v}_3(\sqrt{G_2}-\sqrt{G_2-1})].
\end{eqnarray}
Under the condition of $ G_2 \gg 1 $, the last term in Eq. \eqref{D31} disappears, resulting in $\hat{a}'{}_4 \approx \hat{a}_2$. This indicates that the initially independent particles $\hat{a}'{}_4$ and $\hat{a}_1$ become entangled after AOES \cite{ref-29}. Combining Eq. \eqref{A7}, the Heisenberg expression for $\hat{a}'{}_4$ from Eq. \eqref{D3} is
\begin{eqnarray}
&\hat{x}'{}_{a_4}=\frac{1}{\sqrt{2}}(e^{+r_1}\hat{x}_{a_0}-e^{-r_1}\hat{x}_{v_1})+\sqrt{\frac{G_3-1}{G_3}}(\sqrt{2}e^{-r_2}\hat{x}_{v_3}),\nonumber\\
&\hat{p}'{}_{a_4}=\frac{1}{\sqrt{2}}(e^{-r_1}\hat{p}_{a_0}-e^{+r_1}\hat{p}_{v_1})+\sqrt{\frac{G_3-1}{G_3}}(\sqrt{2}e^{-r_2}\hat{p}_{v_2}).
\label{eq13}
\end{eqnarray}

In order to characterize the performance of AOES, we conducted a Bell test based on ref. \cite{4,Oliver2018}. As shown in Figure \ref{fig3}, Alice, Bob, and Charlie perform two of the above AOES at the same time; one allocates $\hat{a}_1$ and $\hat{a}_4$ respectively to Alice and Charlie, $\hat{a}_2$ and $\hat{a}_3$ to Bob, while the other allocates $\hat{b}_1$ and $\hat{b}_4$ to Alice and Charlie, $\hat{b}_2$ and $\hat{b}_3$ to Bob. Alice, Bob and Charlie are simultaneously executing two identical AOES processes. One process, as described above, results in $\hat{a}_1$ and $\hat{a}'{}_4$ from Eqs. \eqref{A7} and \eqref{eq13}, while the other identical AOES process can produce the entangled $\hat{b}_1$ and $\hat{b}'{}_4$, namely:
\begin{eqnarray}\label{B1}
&\hat{x}_{b_1}=\frac{1}{\sqrt{2}}(e^{+r_1}\hat{x}_{b_0}+e^{-r_1}\hat{x}_{v_4}),
\nonumber\\
&\hat{p}_{b_1}=\frac{1}{\sqrt{2}}(e^{-r_1}\hat{p}_{b_0}+e^{+r_1}\hat{p}_{v_4}),\nonumber\\
&\hat{x}'{}_{b_4}=\frac{1}{\sqrt{2}}(e^{+r_1}\hat{x}_{b_0}-e^{-r_1}\hat{x}_{v_4})+\sqrt{\frac{G_3-1}{G_3}}(\sqrt{2}e^{-r_2}\hat{x}_{v_6}),\nonumber\\
&\hat{p}'{}_{b_4}=\frac{1}{\sqrt{2}}(e^{-r_1}\hat{p}_{b_0}-e^{+r_1}\hat{p}_{v_4})+\sqrt{\frac{G_3-1}{G_3}}(\sqrt{2}e^{-r_2}\hat{p}_{v_5}).
\end{eqnarray}

After the above AOES process, Alice has beam $\hat{b}_1$ (horizontal polarization) and beam $\hat{a}_1$ (vertical polarization), and then combines them using PBS to form two output modes $\hat{a}_h$ and $\hat{a}_v$. Charlie reorganizes the beams $\hat{a}'{}_4$ and $\hat{b}'{}_4$ in the same way to form two output modes $\hat{c}_h$ and $\hat{c}_v$.
Alice and Charlie combined their polarization-separated states at angles $\theta$ and $\vartheta$ using a halfwave plate $(\lambda/2)$ and a polarizing beam splitter (PBS)\cite{Oliver2018}. The resulting states, $\hat{a}_+$, $\hat{a}_-$, $\hat{c}_+$, and $\hat{c}_-$, were measured with homodyne detectors, described by the following transformation
\begin{eqnarray}\label{D7}
&\hat{a}_+(\theta)=\cos{\theta}\hat{a}_h+\sin{\theta}\hat{a}_v,
\nonumber\\
&\hat{a}_-(\theta)=-\sin{\theta}\hat{a}_h+\cos{\theta}\hat{a}_v,\nonumber\\
&\hat{c}_+(\vartheta)=\cos{\vartheta}\hat{c}_h+\sin{\vartheta}\hat{c}_v, 
\nonumber\\
&\hat{c}_-(\vartheta)=-\sin{\vartheta}\hat{c}_h+\cos{\vartheta}\hat{c}_v.
\end{eqnarray}
Their Heisenberg expressions are given by 
\begin{eqnarray}\label{D8}
&\hat{x}^+_a=\cos{\theta}\hat{x}_{b_1}+\sin{\theta}\hat{x}_{a_1},\nonumber\\
&
\hat{p}^+_a=\cos{\theta}\hat{p}_{b_1}+\sin{\theta}\hat{p}_{a_1},\nonumber\\
&\hat{x}^+_c=\cos{\vartheta}\hat{x}'{}_{a_4}+\sin{\vartheta}\hat{x}'{}_{b_4},\nonumber\\
&
\hat{p}^+_c=\cos{\vartheta}\hat{p}'{}_{a_4}+\sin{\vartheta}\hat{p}'{}_{b_4},\nonumber\\
&\hat{x}^-_a=-\sin{\theta}\hat{x}_{b_1}+\cos{\theta}\hat{x}_{a_1},\nonumber\\
&
\hat{p}^-_a=-\sin{\theta}\hat{p}_{b_1}+\cos{\theta}\hat{p}_{a_1},\nonumber\\
&\hat{x}^-_c=-\sin{\vartheta}\hat{x}'{}_{a_4}+\cos{\vartheta}\hat{x}'{}_{b_4},\nonumber\\
&
\hat{p}^-_c=-\sin{\vartheta}\hat{p}'{}_{a_4}+\cos{\vartheta}\hat{p}'{}_{b_4}.
\end{eqnarray}

The correlators of $A_xC_z$ conditional on the outcome Bob is evaluated by using the recorded  photon numbers as 
\begin{eqnarray}\label{B11}
\langle A_xC_z|B^b\rangle &=\frac{\sum_{a,c=\pm1}ac R^{a,c|b}(x,z)}{\sum_{a,c=\pm1} R^{a,c|b}(x,z)}
\end{eqnarray}
where the photon numbers $R^{a,c|b}(x,z)$ of Alice and Charlie conditional on the outcome $b$ of Bob are defined by  
\begin{eqnarray}
\label{B12}
 R^{a,c|b}(x,z)&=\langle R^{a}(x)R^{c}(z)\rangle\nonumber\\&=\langle \hat{a}^\dag_{a}(x)\hat{a}_a(x)\hat{c}^\dag_c(z)\hat{c}_c(z)\rangle
\end{eqnarray}
    for $a,c\in \{+,-\}$. 
    
The correlation functions are defined in terms of photon number measurements, like $\hat{a}^\dagger_i(\theta)\hat{a}_i(\theta)$. This has been recently decomposed into a series of quadrature amplitude measurements on the subsystems and their measurement environment using the equivalence \cite{4}: 
\begin{eqnarray}\label{B13}
\hat{a}^\dagger_a\hat{a}_a&\equiv \hat{a}^\dagger_a\hat{a}_a-\hat{v}^\dagger\hat{v}\nonumber\\
&=\hat{x}_{a}^2+\hat{p} _{a}^2-\hat{x}_{v}^2-\hat{p}_{v}^2,
\end{eqnarray}
where $\hat{v}$ is a vacuum mode such that $\langle\hat{v}^{\dagger}\hat{v}\rangle=0$.
Experimentally, $\hat{x}$ and $\hat{p}$ can be measured using the balanced homodyne technique.
Using the equivalence relation in Eq. \eqref{B13}, the correlation in Eq. \eqref{B12} can be expressed in terms of homodyne quadrature measurements. Assuming Gaussian statistics, all correlations can be reduced to second-order correlations. In this scenario, using $\langle\hat{X}^2\hat{Y}^2\rangle=\langle\hat{X}^2\rangle\langle\hat{Y}^2\rangle+2\langle\hat{X}\hat{Y}\rangle^2$, we have
\begin{eqnarray}\label{B14}
R^{a,c|b}=&2(\langle\hat{x}_a\hat{x}_c\rangle^2+\langle\hat{p}_a\hat{p}_c\rangle^2
+\langle\hat{x}_a\hat{p}_c\rangle^2+\langle\hat{p}_a\hat{x}_c\rangle^2)\nonumber\\
&+V_{a;x}V_{c;x}+
V_{a;p}V_{c;p}+V_{a;p}V_{c;x}\nonumber\\
&+V_{a;x}V_{c;p}-2V_v(V_{a;x}+V_{a;p})\nonumber\\
&-2V_v(V_{a;x}+V_{c;p})+4V^2_v.
\end{eqnarray}
Here, $V^i_{F,k}=\langle\hat{k}_F^2\rangle$ for $\hat{k}\in\{\hat{x},\hat{p}\}$, where $F$ is the mode $a$, $c$, and $V_v$ is the second moment of the vacuum mode.

When $\hat{a}_0$ and $\hat{b}_0$ are in the vacuum state, and under the condition that Bob's result is $b$, further calculations imply that 
\begin{eqnarray}\label{D10}
&\langle\hat{x}_{a_1}^2\rangle=\langle\hat{p}_{a_1}^2\rangle=\langle\hat{x}_{b_1}^2\rangle=\langle\hat{p}_{b_1}^2\rangle=\frac{1}{4}\cosh{2r_1},
\nonumber\\
&\langle(\hat{x}'{}_{a_4})^2\rangle=\langle(\hat{p}'{}_{a_4})^2\rangle=\langle(\hat{x}'{}_{b_4})^2\rangle=\langle(\hat{p}'{}_{b_4})^2\rangle\nonumber\\&\qquad\quad=\frac{1}{4}(\cosh{2r_1}+\frac{G_3-1}{G_3}2e^{-2r_2}),\nonumber\\
&\langle\hat{x}_{a_1}\hat{x}'{}_{a_4}\rangle=-\langle\hat{p}_{a_1}\hat{p}'{}_{a_4}\rangle=\langle\hat{x}_{b_1}\hat{x}'{}_{b_4}\rangle
\nonumber\\
&\qquad\ \ \ \ \ \ =-\langle\hat{p}_{b_1}\hat{p}'{}_{b_4}\rangle=\frac{1}{4}\sinh{2r_1},
\nonumber\\
&\langle\hat{x}_{a_1}\hat{x}_{b_1}\rangle=\langle\hat{x}'{}_{a_4}\hat{x}'{}_{b_4}\rangle=\langle\hat{p}_{a_1}\hat{p}_{b_1}\rangle=\langle\hat{p}'{}_{a_4}\hat{p}'{}_{b_4}\rangle=0,\nonumber\\
&\langle\hat{x}_{b_1}\hat{x}'{}_{a_4}\rangle=\langle\hat{x}_{a_1}\hat{x}'{}_{b_4}\rangle=\langle\hat{p}_{b_1}\hat{p}'{}_{a_4}\rangle=\langle\hat{p}_{a_1}\hat{p}'{}_{b_4}\rangle=0,\nonumber\\
&\langle\hat{p}_{a_1}\hat{x}'{}_{a_4}\rangle=\langle\hat{p}_{a_1}\hat{x}'{}_{b_4}\rangle=\langle\hat{p}_{b_1}\hat{x}'{}_{a_4}\rangle=\langle\hat{p}_{b_1}\hat{x}'{}_{b_4}\rangle=0,
\nonumber
\\
&\langle\hat{x}_{b_1}\hat{p}'{}_{b_4}\rangle=\langle\hat{x}_{a_1}\hat{p}'{}_{b_4}\rangle=\langle\hat{x}_{a_1}\hat{p}'{}_{b_4}\rangle=\langle\hat{x}_{b_1}\hat{p}'{}_{a_4}\rangle=0.
\end{eqnarray}
By combining Eqs. \eqref{D8} and \eqref{D10}, the mean and variance of some can be calculated under the condition that Bob's result is $b$ and Alice and Charlie's measurement results are $+$:
\begin{eqnarray}\label{D11}
&\langle\hat{x}_a^+\hat{p}_c^+\rangle=\langle\hat{p}_a^+\hat{x}_c^+\rangle=0,\nonumber\\
&\langle\hat{x}_a^+\hat{x}_c^+\rangle=-\langle\hat{p}_a^+\hat{p}_c^+\rangle=\frac{1}{4}\sin(\theta+\vartheta)\sinh{2r_1},\nonumber\\
&\langle(\hat{x}_a^+)^2\rangle=\langle(\hat{p}_a^+)^2\rangle=\frac{1}{4}(\Gamma_1+1),\nonumber\\
&\langle(\hat{x}_c^+)^2\rangle=\langle(\hat{p}_c^+)^2\rangle=\frac{1}{4}(\Gamma_2+1).
\end{eqnarray}
where $\Gamma_i$ are defined by $\Gamma_1=\cosh{2r_1}-1$ and $\Gamma_2=\cosh{2r_1}+\frac{2G_3-2}{G_3}e^{-2r_2}-1$.

Further, by bringing Eqs. \eqref{D11} into Eq. \eqref{B14}, we can get that the photon number correlation function conditional on Bob's outcome $b$ when the measurements are both $+$ is:
\begin{eqnarray}\label{D12}
R^{+,+|b}&=2(\langle\hat{x}^+_a\hat{x}^+_c\rangle^2+\langle\hat{p}^+_a\hat{p}^+_c\rangle^2+\langle\hat{x}^+_a\hat{p}^+_c\rangle^2+\langle\hat{p}^+_a\hat{x}^+_c\rangle^2)\nonumber\\
&\quad+(\langle(\hat{x}^+_a)^2\rangle+\langle(\hat{p}^+_a)^2\rangle-\frac{1}{2})\nonumber\\
&\quad\times(\langle(\hat{x}^+_c)^2\rangle+\langle(\hat{p}^+_c)^2\rangle-\frac{1}{2})\nonumber\\
&=\frac{1}{4}(\sin^2(\theta+\vartheta)\sinh^2{2r_1}+\Gamma_1\Gamma_2).
\end{eqnarray}

Similarly, by using the Eqs. \eqref{D8} and \eqref{D10}, we can obtain the mean and variance of Alice and Charlie's position and momentum combinations conditional on Bob's outcome $b$ when both of their measurements are -:
\begin{eqnarray}\label{D13}
&\langle\hat{x}_a^-\hat{p}_c^-\rangle=\langle\hat{p}_a^+\hat{x}_c^+\rangle=0,\nonumber\\
&\langle\hat{x}_a^-\hat{x}_c^-\rangle=-\langle\hat{p}_a^-\hat{p}_c^-\rangle=-\frac{1}{4}\sin(\theta+\vartheta)\sinh{2r_1},\nonumber\\
&\langle(\hat{x}_a^-)^2\rangle=\langle(\hat{p}_a^-)^2\rangle=\frac{1}{4}(\Gamma_1+1),\nonumber\\
&\langle(\hat{x}_c^-)^2\rangle=\langle(\hat{p}_c^-)^2\rangle=\frac{1}{4}(\Gamma_2+1).
\end{eqnarray}
By substituting Eqs. \eqref{D13} into Eq. \eqref{B14}, we obtain the photon number correlation function conditional on Bob's outcome $b$ when both measurements are $-$:
\begin{eqnarray}\label{D14}
R^{-,-|b}&=2(\langle\hat{x}^-_a\hat{x}^-_c\rangle^2+\langle\hat{p}^-_a\hat{p}^-_c\rangle^2+\langle\hat{x}^-_a\hat{p}^-_c\rangle^2+\langle\hat{p}^-_a\hat{x}^-_c\rangle^2)\nonumber\\
&\quad+(\langle(\hat{x}^-_a)^2\rangle+\langle(\hat{p}^-_a)^2\rangle-\frac{1}{2})\nonumber\\
&\quad\times(\langle(\hat{x}^-_c)^2\rangle+\langle(\hat{p}^-_c)^2\rangle-\frac{1}{2})\nonumber\\
&=\frac{1}{4}(\sin^2(\theta+\vartheta)\sinh^2{2r_1}+\Gamma_1\Gamma_2).
\end{eqnarray}

Given Bob's outcome $b$, when Alice measures $+$ and Charlie measures $-$, we can combine Eqs. \eqref{D8} and \eqref{D10} to obtain:
\begin{eqnarray}\label{D15}
&\langle\hat{x}_a^+\hat{p}_c^-\rangle=\langle\hat{p}_a^+\hat{x}_c^-\rangle=0,\nonumber\\
&\langle\hat{x}_a^+\hat{x}_c^-\rangle=-\langle\hat{p}_a^+\hat{p}_c^-\rangle=\frac{1}{4}\cos(\theta+\vartheta)\sinh{2r_1},\nonumber\\
&\langle(\hat{x}_a^+)^2\rangle=\langle(\hat{p}_a^+)^2\rangle=\frac{1}{4}(\Gamma_1+1),\nonumber\\
&\langle(\hat{x}_c^-)^2\rangle=\langle(\hat{p}_c^-)^2\rangle=\frac{1}{4}(\Gamma_2+1).
\end{eqnarray}
Further, by bringing Eqs.\eqref{D15} into Eq. \eqref{B14}, we can get the quantum correlation function conditional on Bob's outcome $b$ for Alice's measurement of $+$ and Charlie's measurement of $-$:
\begin{eqnarray}\label{B16}
R^{+,-|b}&=2(\langle\hat{x}^+_a\hat{x}^-
_c\rangle^2+\langle\hat{p}^+_a\hat{p}^-_c\rangle^2+\langle\hat{x}^+_a\hat{p}^-_c\rangle^2+\langle\hat{p}^+_a\hat{x}^-_c\rangle^2)\nonumber\\
&\quad+(\langle(\hat{x}^+_a)^2\rangle+\langle(\hat{p}^+_a)^2\rangle-\frac{1}{2})\nonumber\\
&\quad
\times(\langle(\hat{x}^-_c)^2\rangle+\langle(\hat{p}^-_c)^2\rangle-\frac{1}{2})\nonumber\\
&=\frac{1}{4}(\cos^2(\theta+\vartheta)\sinh^2{2r_1}+\Gamma_1\Gamma_2).
\end{eqnarray}

Given Bob's outcome $b$, when Alice measures $-$ and Charlie measures $+$, we can combine Eqs. \eqref{D8} and \eqref{D10} to obtain:
\begin{eqnarray}\label{D17}
&\langle\hat{x}_a^-\hat{p}_c^+\rangle=\langle\hat{p}_a^-\hat{x}_c^+\rangle=0,\nonumber\\
&\langle\hat{x}_a^-\hat{x}_c^+\rangle=-\langle\hat{p}_a^-\hat{p}_c^+\rangle=\frac{1}{4}\cos(\theta+\vartheta)\sinh{2r_1},\nonumber\\
&\langle(\hat{x}_a^-)^2\rangle=\langle(\hat{p}_a^-)^2\rangle=\frac{1}{4}(\Gamma_1+1),\nonumber\\
&\langle(\hat{x}_c^+)^2\rangle=\langle(\hat{p}_c^+)^2\rangle=\frac{1}{4}(\Gamma_2+1).
\end{eqnarray}
By bringing Eqs. \eqref{D17} into Eq. \eqref{B14}, we can get the quantum correlation function conditional on Bob's outcome $b$ for Alice's measurement of $-$ and Charlie's measurement of $+$:
\begin{eqnarray}
\label{D18}
R^{-,+|b}&
=2(\langle\hat{x}^-
_a\hat{x}^+_c\rangle^2+\langle\hat{p}^-_a\hat{p}^+_c\rangle^2+\langle\hat{x}^-_a\hat{p}^+_c\rangle^2+\langle\hat{p}^-_a\hat{x}^+_c\rangle^2)\nonumber\\
&\quad+(\langle(\hat{x}^-_a)^2\rangle+\langle(\hat{p}^-_a)^2\rangle-\frac{1}{2})\nonumber\\
&\quad\times (\langle(\hat{x}^+_c)^2\rangle+\langle(\hat{p}^+_c)^2\rangle-\frac{1}{2})\nonumber\\
&=\frac{1}{4}(\cos(\theta+\vartheta)^2\sinh^2{2r_1}+\Gamma_1\Gamma_2).
\end{eqnarray}

 By bringing Eqs. \eqref{D12},  \eqref{D14},  \eqref{B16}, and  \eqref{D18} into Eq. \eqref{B11}, we get 
\begin{eqnarray}\label{Eq20}
\langle A_xC_z|B^b\rangle=\frac{-\cos{2(\theta+\vartheta)}\sinh^2{2r_1}}{\sinh^2{2r_1}+2\Gamma_1\Gamma_2}.
\end{eqnarray}
Let $x\in\{0,1\}$ correspond to the angles of Alice's polarizer $\theta =3\pi/8$ and $\theta=\pi/8$ , and $z\in\{0,1\}$ correspond to the angles of Charlie's polarizer $\vartheta=\pi /4$ and $\vartheta=0$, respectively. And when $G_3=8 \gg 1$, $\mathcal{B}_{b}$ in inequality \eqref{Eq2b} can be expressed as
\begin{eqnarray}\label{Eq21}
\mathcal{B}_{b}=\frac{2\sqrt{2}\sinh^2{2r_1}}{\sinh^2{2r_1}+2\Gamma_1\Gamma_2}.
\end{eqnarray}

\begin{figure}
\begin{center}
\includegraphics[scale=0.6]{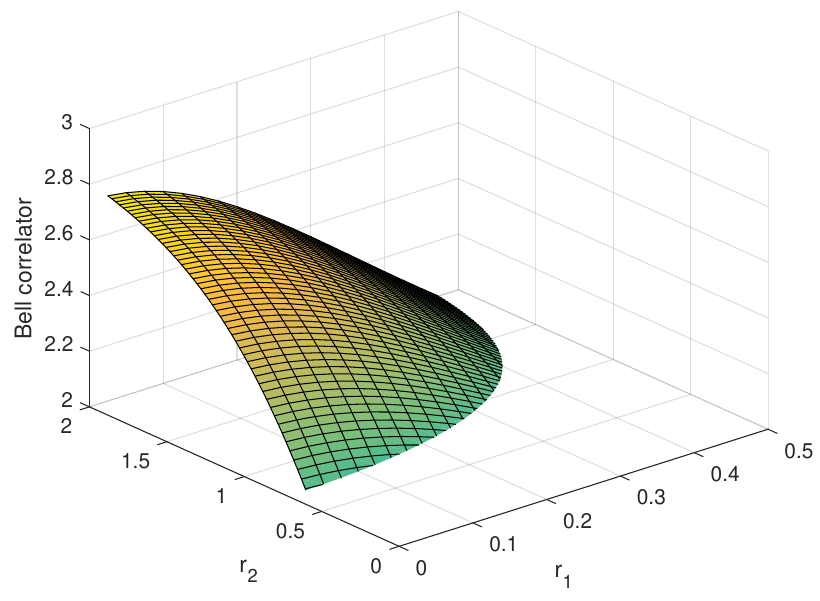}
\end{center}
\caption {The Bell correlators via  compression parameters $r_1$ and $r_2$. }
\label{fig4}
\end{figure}

As shown in Figure \ref{fig4}, there are suitable compression parameters $r_1$ and $r_2$ for the system such that the quantum correlations violate the Bell inequality \eqref{Eq2b}.

\section{Discussion}

We have developed an all-optical method to verify the one-way network nonlocality of tripartite continuous variable (CV) quantum networks. This method can be further generalized to accommodate long-chain CV networks composed of multiple CV bipartite entangled states, although this requires extensive evaluations \cite{Luo2024}. Additionally, one may explore the extension of our approach to star-type networks. Both tripartite and star-type networks can be certified through multipartite experiments, but they also lend themselves to decomposition into subsets of tripartite subnetworks. This characteristic enables a step-by-step verification of nonlocality, facilitating a more manageable experimental approach.

In contrast to discrete variable quantum networks, where conducting Bell tests for multiple sources is relatively straightforward \cite{SBBP,Huang2022,Gu2023,Wang2023,Mao2024}, verifying nonlocality in CV experiments presents significant challenges. The current protocol, as illustrated in Figure \ref{fig4}, necessitates a large compression parameter, which complicates the implementation of the experiment. Overall, our work has provided a comprehensive all-optical experimental design aimed specifically at verifying the one-way network nonlocality of tripartite CV quantum networks, contributing to the understanding and application of quantum resources in this domain.

\section*{Supporting Information}

Supporting Information is available from the Wiley Online Library. 

\section*{Acknowledgements}

This work was supported by the National Natural Science Foundation of China (Nos. 62172341, 12204386,12405024), Sichuan Natural Science Foundation (Nos.2024NSFSC1375,2024NSFSC1365), and Interdisciplinary Research of Southwest Jiaotong University China (No.2682022KJ004).

\section*{Conflict of Interest} 

The authors declare no other conflict of interest.

\section*{Author Contributions} 

M.X.L. conducted the research. All authors wrote and reviewed the manuscript.

\section*{Data Availability Statement} 

This is no data generated in research.

\end{document}